\documentclass[a4paper]{ISOStyle}

\usepackage{graphics}
\newcommand{\ddd}{\,\mathrm{d}}
\newcommand{\figdir}{./}

\title{The LARI Method for ISO-CAM/PHOT Data Reduction and Analysis}

\author{
Carlo Lari\inst{1}\thanks{e-mail: lari@ira.bo.cnr.it}
  \and
Mattia Vaccari\inst{2,3}
  \and
Giulia Rodighiero\inst{2}
  \and
Dario Fadda\inst{4}
  \and
Carlotta Gruppioni\inst{5,6}
  \and
Francesca Pozzi\inst{7}
  \and
Alberto Franceschini\inst{2}
  \and
Gianni Zamorani\inst{6}
} 

\institute{
Institute of Radio Astronomy, CNR, Via Gobetti 101, I-40122, Bologna, Italy
  \and
Department of Astronomy, University of Padova, Vicolo dell'Osservatorio 5, I-35122, Padova, Italy
  \and
CISAS ``G.\ Colombo'', University of Padova, Via Venezia 15, I-35131, Padova, Italy
  \and
Instituto de Astrof\'isica de Canarias, Via Lactea S/N, E-38205, La Laguna, Tenerife, Spain
  \and
Padova Astronomical Observatory, INAF, Vicolo dell'Osservatorio 2, I-35122, Padova, Italy
  \and
Bologna Astronomical Observatory, INAF, Via Ranzani 1, I-40127 Bologna, Italy
  \and
Department of Astronomy, University of Bologna, Via Ranzani 1, I-40127, Bologna, Italy
}

\begin{document}
\maketitle
\begin{abstract}
The techniques and software tools developed for the reduction and analysis of
ISO-CAM/PHOT data with the LARI method are presented.
The method, designed for the detection of faint sources in ISO raster
observations, is based on the assumption of the existence of two
different time scales in the detectors' transient behaviour, accounting either
for fast or slow detectors' response.

The specifically developed IDL software includes: a reduction pipeline
performing basic operations such as deglitching and background determination;
the fitting procedures proper, modelling the time history of individual pixels
and detecting any flux excess with respect to the local background ascribable
to potential sources; mapping, source extraction and flux estimation
procedures; simulation procedures allowing one to estimate the errors arising
from different instrumental and reduction effects.
Moreover, an easy-to-use graphical user interface allows one to quickly browse
the data and carry out the substantial amount of interactive analysis required
when the automatic fit fails and to check the reliability of detected sources.

This method provides source lists of great reliability and completeness
and an outstanding photometric accuracy, particularly at low redundancy levels,
where the reliability of ISO-CAM/PHOT source lists, even at moderately bright
flux levels, has been a long-standing issue.

In this work a description of the techniques and software tools that have been
developed is given, alongside with some highlights from the results obtained
thanks to their application to different fields.
%The completeness, reliability and photometric accuracy of the resulting
%source lists at different flux and raster redundancy levels are then discussed
%through simulations.
%
\end{abstract}

% NB: LONG denotes suggestions to write a (future) longer, self-consistent paper

\section{Introduction}
All data gathered by the ISO satellite, and particularly those from
the its two cameras, ISO-CAM and ISO-PHOT (hereafter simply CAM and PHOT),
are very difficult to reduce, both due to the strong transient behaviour
of the cryogenically cooled detectors, e.g. after a change in the incident
flux (\cite{Coulais2000}), and to the frequent and severe cosmic ray
impacts yielding qualitatively different effects (known as common glitches,
faders, dippers, drop-outs and others, \cite{Claret1998}).
% LONG !!! add figures for the four cases quoted above !!!

While it was variously demonstrated that it is possible, at least to a certain extent,
%(see e.g. Technical Reports by Coulais et al. and Lari)
%Quote Siguenza's poster by Coulais about the numerical uselessness of his model
to satisfactorily describe the satellite's different detectors' behaviour
adopting some physical model, the large number of readouts
involved in raster observations and the peculiar nature and strength of noise
patterns also require efficient and robust algorithms to be developed
so as to make the actual data reduction undertaking feasible in a
nearly-automatic way.

A number of data reduction methods has thus been developed and tested,
mostly on CAM deep fields (e.g. the PRETI method by \cite{Starck1999}
and the Triple Beam Switch method by \cite{Desert1999}).
Unfortunately, such methods proved useless for all PHOT data or on
CAM shallower fields, leading to a high number of false detections
and severe incompleteness. Besides, these methods suffered from the lack
of an efficient way to interactively check the quality of the data reduction
when needed.

The LARI method, originally developed as an answer to the problems posed
by the reduction of the observatons carried out as part of the ELAIS survey
(\cite{Oliver2000}) and first presented in \cite{Lari2001}, has been devised
to overcome these difficulties and provide a fully-interactive technique
for the reduction and analysis of CAM/PHOT raster observations
at all flux levels, particularly suited for the detection of faint
sources and thus for the full exploitation of the scientific potential
of the ISO archive.
\section{The Model\label{model.sec}}
The LARI method describes the sequence of readouts, or time history,
of each pixel of CAM/PHOT detectors in terms of a mathematical model
for the charge release towards the contacts.
Such a model is based on the assumption of the existence, in each pixel,
of two charge reservoirs, a short-lived one $Q_b$ (breve) and a
long-lived one $Q_l$ (lunga), evolving independently with a different
time constant and fed by both the photon flux and the cosmic rays.
Such a model is fully conservative, and thus the observed signal $S$ is
related to the incident photon flux $I$ and to the accumulated charges
$Q_b$ and $Q_l$ by the
\begin{equation}
S = I - \frac{\ddd Q_{tot}}{\ddd t} = I - \frac{\ddd Q_b}{\ddd t} - \frac{\ddd Q_l}{\ddd t}~,
\end{equation}
where the evolution of these two quantities is governed by the same
differential equation, albeit with a different efficiency $e_i$
and time constant $a_i$
\begin{equation}
\frac{\ddd Q_i}{\ddd t} = e_i\,I - a_i\,Q_{i}^2~~~~~\mathrm{where}~~~i=b,l~,
\end{equation}
so that
\begin{equation}
S = (1-e_b-e_l)\,I + a_b\,Q_b^2 + a_l\,Q_l^2~.
\end{equation}
The values of the parameters $e_i$ and $a_i$ are estimated from the data and
are constant for a given detector, apart from the scaling of $a_i$ for the
exposure time and the average signal level along the pixel time history,
which is governed by the
\begin{equation}
a_i = \frac{t}{t_0}\,\sqrt{\frac{S}{S_0}}~a_{i,0}~,
\end{equation}
where $a_{i,0}$ is the value of $a_i$ relative to a reference exposure time
$t_0$ and average signal level $S_0$.
The model for the charge release, however, is exactly the same for
CAM and PHOT detectors.
%!!! Define an efficiency for breve and lunga reservoirs and a time constant
%for given ``standard'' flux level and exposure time: tabulate the values
%for CAM LW and PHOT C-100 detectors !!!

In practice, an additive offset signal attributable to thermal dark current
(which is not accounted for in CIA/PIA pipeline dark current subtraction),
is added to both $S$ and $I$ in the equation above when it is estimated to be
important, i.e. when the deepest dippers' depth exceeds 10\% of the background
level.

The glitches (i.e. the effects of cosmic ray impacts on time history)
are identified through filtering of the time history and modelled as
discontinuities in the charge release, leaving as free parameters the charges
at the beginning of the time history and at the peaks of glitches.
%Glitches from nearby pixels are also considered when their height is
%substantially (e.g. 20 times) higher than the chosen threshold
%strongest glitches from the pixel + those from nearby pixels

\begin{figure}[!ht]
\begin{center}
\resizebox{0.9\columnwidth}{0.49\columnwidth}{\includegraphics{\figdir/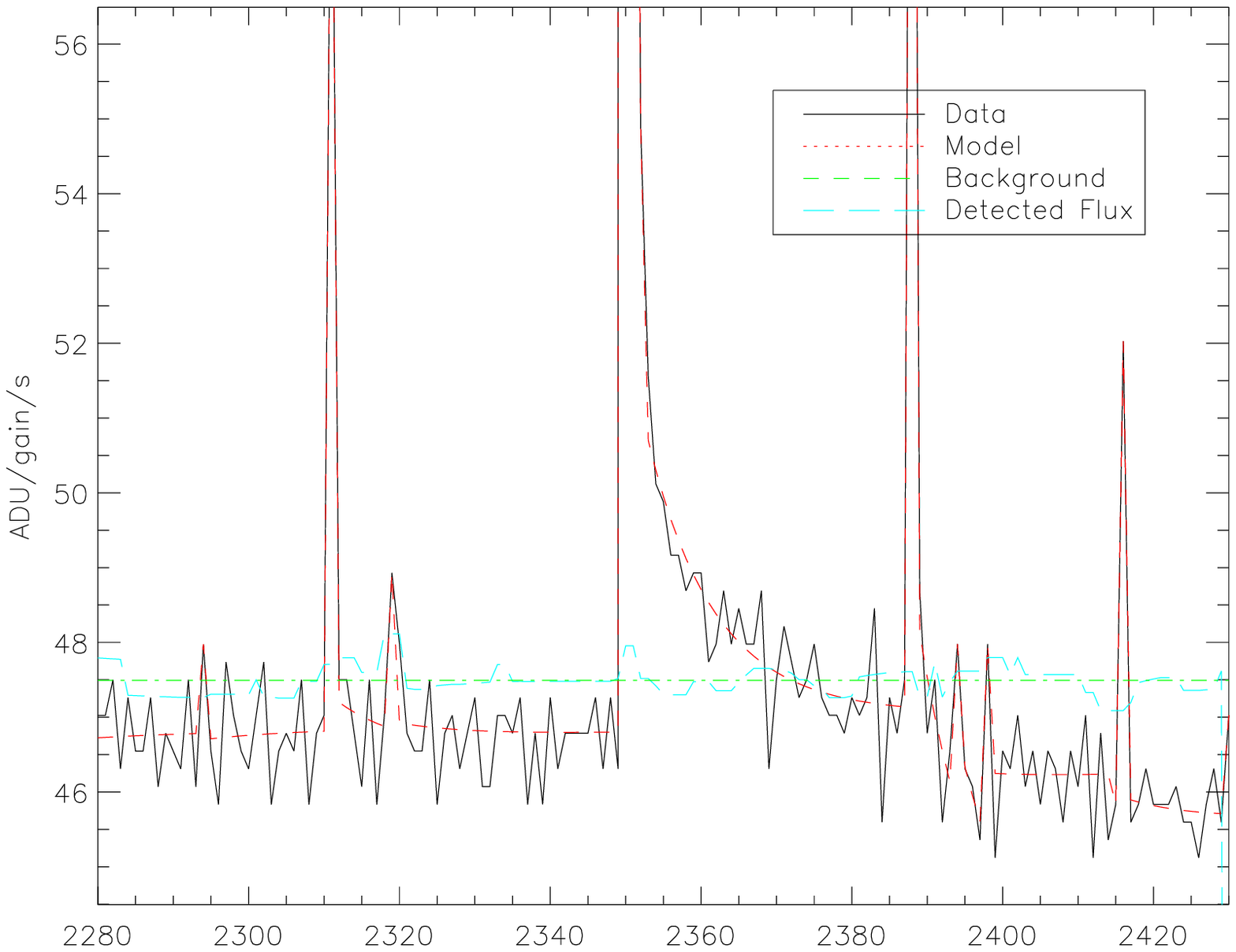}}
\\\textbf{a) Fader}\vspace{0.25cm}\\
\resizebox{0.9\columnwidth}{0.49\columnwidth}{\includegraphics{\figdir/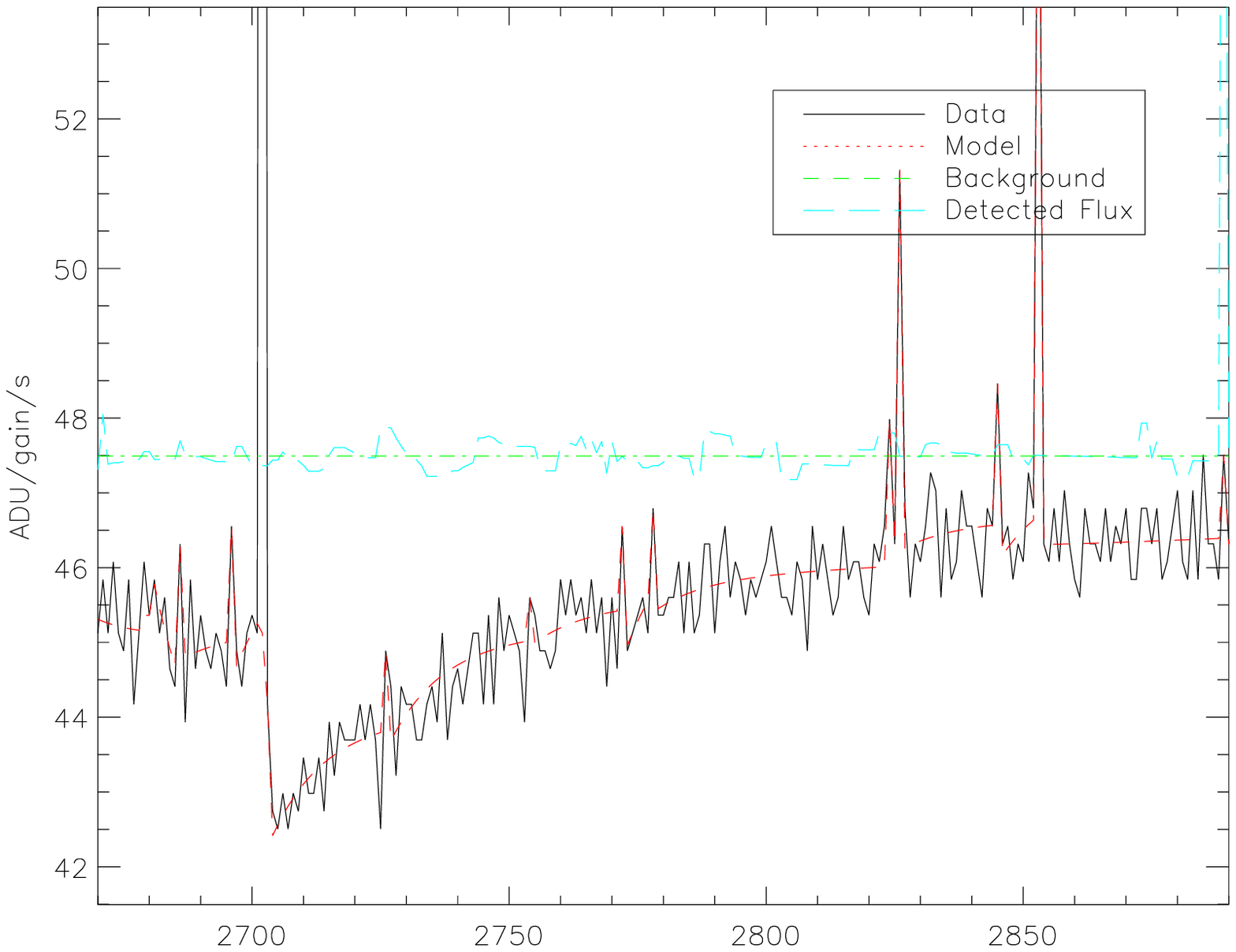}}
\\\textbf{b) Dipper}\vspace{0.25cm}\\
\resizebox{0.9\columnwidth}{0.49\columnwidth}{\includegraphics{\figdir/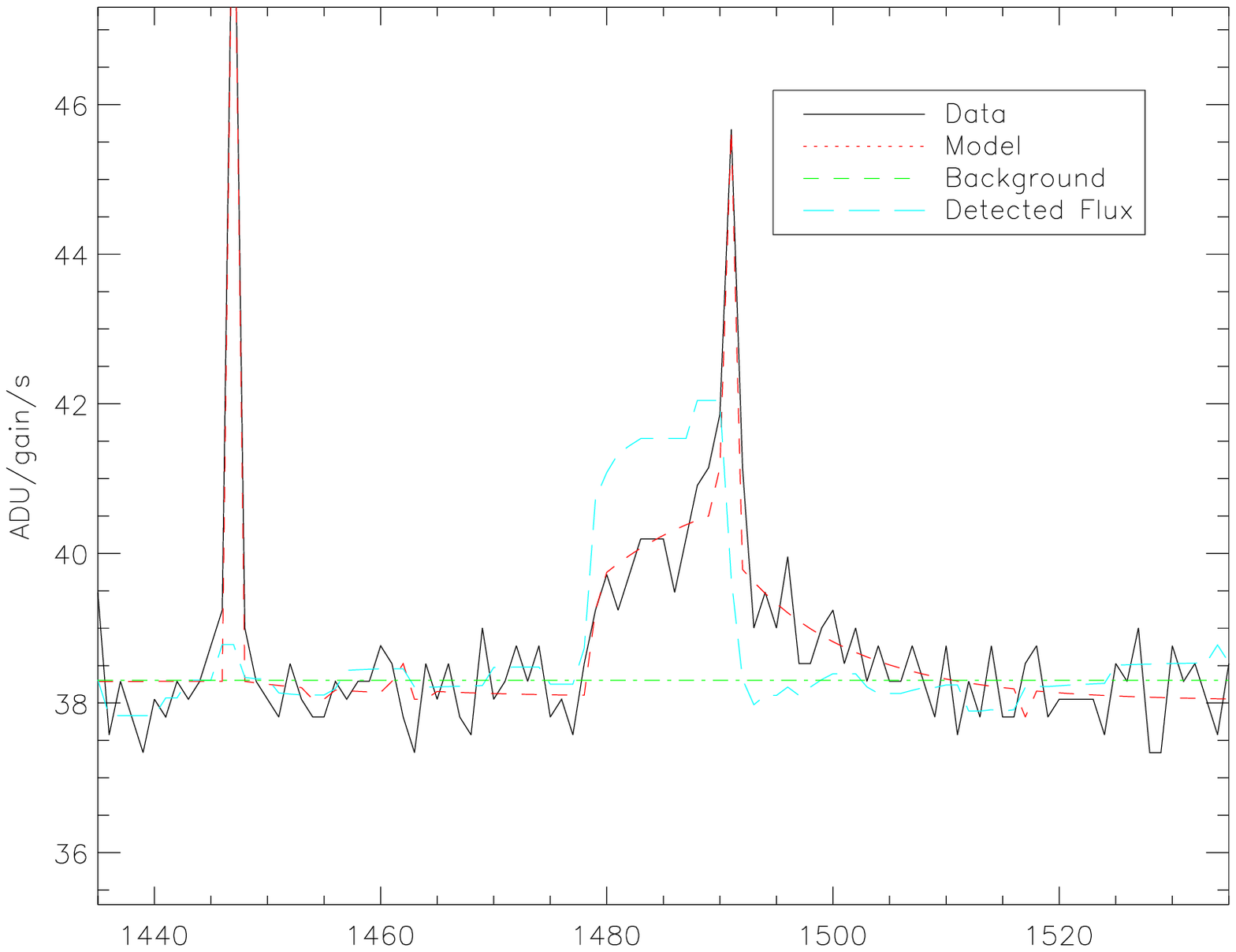}}
\\\textbf{c) Bright source}\vspace{0.25cm}\\
\resizebox{0.9\columnwidth}{0.49\columnwidth}{\includegraphics{\figdir/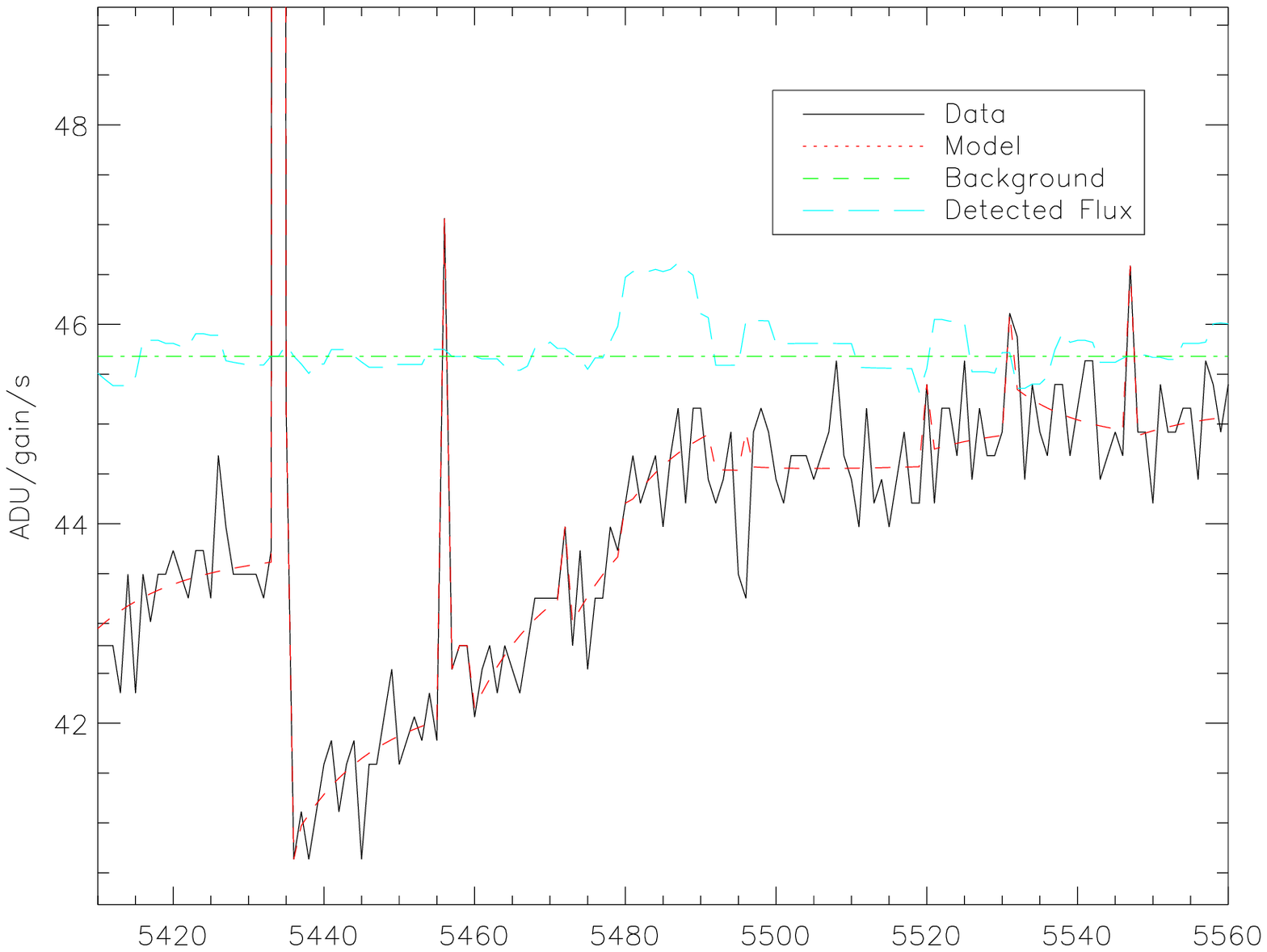}}
\\\textbf{d) Faint source}\vspace{0.25cm}\\
\caption{Different troublesome situations in CAM pixel time histories:
a) Recovery of stabilization background level after a fader b) Recovery of stabilization background level after a dipper c) Detection of a bright source hidden by a strong common glitch d) Faint source hidden by the recovery of the stabilization background level after a dipper.}
\label{fits.fig}
\end{center}
\end{figure}

Iteration of the fitting procedure is interrupted when either a satisfactory
data-model rms deviation is achieved or the maximum number of allowed
iterations is reached.
In Figure~\ref{fits.fig} it is shown how a successful fit is thus able to
recover useful information (specifically, source fluxes) from otherwise
troublesome parts of the pixel time history.
%
% \begin{figure}[!ht]
% \begin{center}
% \begin{minipage}{4.25cm}
% \centering
% \resizebox{\textwidth}{0.6\textwidth}{\includegraphics{\figdir/LariC_1.eps}}
% \\\textbf{a) Fader}
% \end{minipage}
% \hspace{-0.3cm}
% \begin{minipage}{4.25cm}
% \centering
% \resizebox{\textwidth}{0.6\textwidth}{\includegraphics{\figdir/LariC_2.eps}}
% \\\textbf{b) Dipper}
% \end{minipage}
% \end{center}
% \vspace{-0.2cm}
% \begin{center}
% \begin{minipage}{4.25cm}
% \centering
% \resizebox{\textwidth}{0.6\textwidth}{\includegraphics{\figdir/LariC_3.eps}}
% \\\textbf{c) Bright source}
% \end{minipage}
% \hspace{-0.3cm}
% \begin{minipage}{4.25cm}
% \centering
% \resizebox{\textwidth}{0.6\textwidth}{\includegraphics{\figdir/LariC_4.eps}}
% \\\textbf{d) Faint source}
% \end{minipage}
% \caption{Different troublesome situations in ISO-CAM pixel time history:
% a) Fader b) Dipper c) Bright source with a strong common glitch d) Faint source over a dipper.}
% \label{fits.fig}
% \end{center}
% \end{figure}
%
\section{The Method}
The reduction pipeline consists of the following steps:
\begin{itemize}
\item
PHOT ramps' linearization (following \cite{Rodighiero2001}) and data
smoothing (i.e. median averaging over a suitable number of readouts)
% usually 32, i.e. half a ramp (made of 64*1/32 second readouts), i.e. 1 second
\item
Standard CIA/PIA \texttt{raster} structure and data-reduction-dedicated 
\texttt{liscio} structure building
%(Readouts expressed in ADU/gain/s)
\item
Dark current subtraction, stabilization (or global, as opposed to local,
see below) background level estimation, bright sources' and glitches'
identification 
\item
Time history fitting procedure and interactive "repair" on fitting failures
\item
Interactive checks on sources detected in time history
\item
Flat-fielding, mapping (i.e. projection of fluxes measured along the pixel time
histories onto a sky map), and source extraction (using DAOPHOT's
\texttt{find}, particularly suited for the detection of point-like sources,
as implemented in IDL Astronomy User's Library)
\item
Interactive checks on sources detected on sky maps and back-projected on
pixel time histories
\item
Source flux autosimulation
\end{itemize}
The fitting procedure describes the time history of individual pixels according
to the mathematical model seen in Section~\ref{model.sec}, allowing the
determination of the breve and lunga charge levels, the local background
(i.e. the signal to be expected on the basis of the previously accumulated
charges if only the stabilization background flux were hitting the detector)
and the flux excess ascribable to potential sources at any given readout).
%
%The fitting procedure allows the determination of the charges...
%
%Describe the nature of ``repair'', glitch addition or subtraction,
%background determination \ldots
%
%A new mapping technique with respect to \cite{Lari2001} is now employed
%to overcome the severe PSF undersampling, consisting in a 3x subsampling
%with respect to the original pixel size.
%
% at any given time, rec and unrec maps are produced... only the former ones,
% however, are used for the following data analysis.
%
% All projections (time history --> sky, rasters --> mosaic, simulated/real
% sources --> expected flux (autosimulation) were carried out using the
% \texttt{projection} C++ code
% included in CIA (for CAM, field distortion maps were provided by Aussel!!!)
% Write Aussel and ask for report about projection code!!!

The so-called autosimulation procedure for source flux estimation
accounts for mapping effects in the determination of the total flux of
detected sources through the following steps:
%(which can be severe, owing to the large PSF undersampling
%implied by the large pixel size)
%
% transients can only be corrected through simulations or flux calibration
% In both cases one may think of determnining a flux-dependent correction!!!
%
\begin{itemize}
\item First guess of source total flux, based on its observed peak flux on the map
\item Back-projection of source from sky map on pixel time histories, i.e.
construction of theoretical pixel time histories
\item Projection of theoretical pixel time histories onto a sky map, i.e. construction
of a theoretical sky map
\item Determination of theoretical peak flux on theroretical sky map
\item Source firts-guess total flux correction based on observed / theoretical peak flux ratio
\end{itemize}
%
% for extended sources this procedure doesn't work, and aperture photometry
% is carried out instead!
%
Other factors affecting the source flux estimates, namely those arising
from the detectors' transient behaviour and possible systematic deviations
from nominal sensitivities as well as from the reduction technique,
are then evaluated through simulations and absolute flux calibration.

Once the reduction of all rasters of interest has been completed according
to the recipe above, one can determine the necessary corrections to nominal
astrometry (e.g. performing a cross-correlation analysis between the list
of detected sources and a suitable external catalogue)
%(offsets)
and project nearby or repeated fields onto a common mosaic sky map, on which
source extraction, autosimulation and interactive checks can furtherly be
performed so as to increase the quality of the reduction through cross-checks
of detected sources on different rasters, thus partly overcoming the severe
problems at their boundaries.
\section{The Software}
\begin{figure}[!ht]
\begin{center}
\centering
\resizebox{0.95\columnwidth}{!}{\includegraphics{\figdir/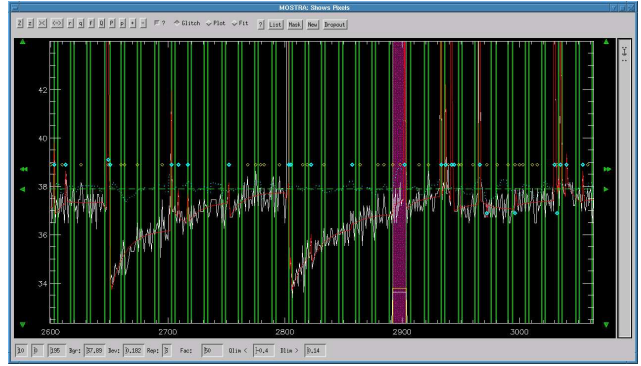}}
\caption{A screenshot of the IDL widget-based Graphical User Interface
used to carry out interactive analysis.}
\label{gui.fig}
\end{center}
\end{figure}

The method relies on CIA (\cite{CIA2001}) and PIA (\cite{PIA1999}) for basic
raw data reading and manipulation and on home-made IDL routines for the data
reduction proper.
The massive necessary amount of interactive analysis is carried out with an
easy-to-use graphical user interface, a screenshot of which is shown in
Figure~\ref{gui.fig}, which allows any kind of visual check and interactive
``repair'' which may be necessary.
\section{Results / Work in Progress}
All parameters indicating the goodness of the reduction
(reliability, completeness, astrometric and photometric accuracy)
are heavily dependent on the adopted observing parameters as well as
on the thresholds chosen in the interactive ``repair", and thus can
only be evaluated through simulations.

\begin{figure}[!ht]
\begin{center}
\begin{minipage}{0.95\columnwidth}
\centering
\resizebox{\textwidth}{!}{\includegraphics{\figdir/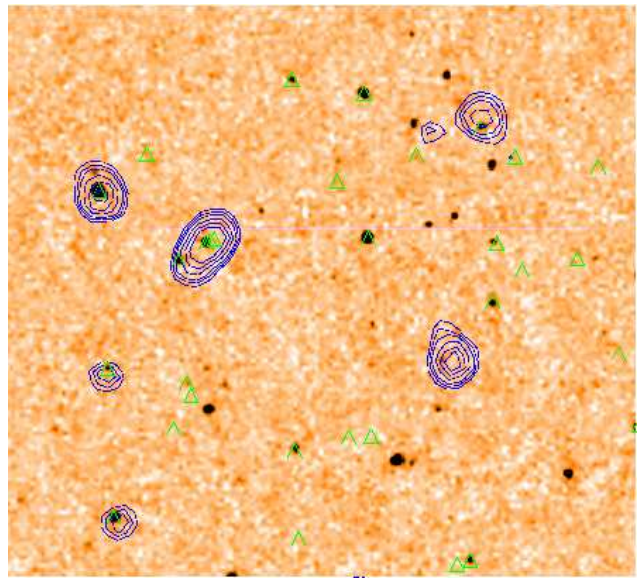}}
\end{minipage}
\caption{
$15^\prime \times 15^\prime$ Lockman Hole 15 $\mu$m map, with overlaid
90 $\mu$m contours (blue lines) and radio sources (green triangles, from
de Ruiter et al. 1997)
}
\label{lhs.fig}
\end{center}
\end{figure}

While the catalogue resulting from the reduction of the first portion
of ELAIS 15 $\mu$m data (\cite{Lari2001}) is successfully being used
for different purposes (see e.g. \cite{Gruppioni2002} and \cite{Matute2002}),
and the reduction of several different fields has already been completed,
simulations and accurate photometric calibration are still being
carried out, so that it is not presently possible to show detailed results.
A list of the different projects being carried out includes:
\begin{itemize}
\item
ELAIS 15 $\mu$m and 90 $\mu$m fields (\cite{Vaccari2002})
\item
Lockman Hole Shallow (LHS) and Deep (LHD) 15 $\mu$m and 90 $\mu$m fields
(\cite{Fadda2002} and \cite{Rodighiero2002}, see also Figure~\ref{lhs.fig})
\item
Hubble Deep Field North and South (HDFs) 7 $\mu$m and 15 $\mu$m fields
\item
A few nearby galaxy cluster 7 $\mu$m and 15 $\mu$m fields
\end{itemize}
while highlights from the expected results can thus be summarized:
%(See also \cite{Fadda2002}, \cite{Rodighiero2002} and \cite{Vaccari2002})
%
\begin{itemize}
\item
A catalogue of around 2000 15 $\mu$m sources in the 0.3-100 mJy flux range
from LHS and ELAIS fields
%for source counts, multi-$\lambda$ identifications and statistical studies
\item
Catalogues of the uttermost quality in smaller, cosmologically relevant fields
such as the LHD and the HDFs
\item
Flux-level-dependent photometric calibration based on predicted stellar
fluxes (CAM) or on internal/external calibrators' reduction (PHOT)
\item
Unambiguous comparison of LARI fluxes with those obtained with different
methods, e.g. on HDFs
\end{itemize}
%
%\begin{center}
%\begin{minipage}{3cm}
%\centering
%\resizebox{0.9\textwidth}{!}{\includegraphics{HDFS_mos_mapx_gaia.eps}}
%{\large\textbf{HDFS NPIX map at 15 $\mu$m}}}
%\end{minipage}
%%\hspace{0.5cm}
%\begin{minipage}{3cm}
%\centering
%\resizebox{0.9\textwidth}{!}{\includegraphics{HDFS_mosaic_gaia.eps}}
%{\large\textbf{HDFS map at 15 $\mu$m}}}
%\end{minipage}
%\end{center}
%
%\begin{wraptable}{r}{4cm}
%\begin{center}
%\begin{large}
%\begin{tabular}{c|ccc|c}
%   &  S1 &  N1 &  N2 & Total \\
%\hline
%PA & 189 & 129 & 141 &  459  \\
%FA-I & $\simeq$ 450 & $\simeq$ 350 & $\simeq$ 400 & $\simeq$ 1200  \\
%FA-II & $\simeq$ 700 & $\simeq$ 550 & $\simeq$  550 & $\simeq$ 1800\\
%\end{tabular}
%\end{large}
%\end{center}
%\end{wraptable}
%
%Number of sources detected in ELAIS 15 $\mu$m (LW3) fields by the
%Preliminary Analysis (Oliver et al. 2001, MNRAS, ?, ?),
%Final Analysis I (Lari et al. 2001) and
%Final Analysis II (See Vaccari's Talk).
%
%While a direct comparison of the fluxes obtained with different methods
%on deep fields is still in progress, preliminary results seem to
%show a good (mostly at the 25\% level) agreement. ?
%
\section{Conclusions}
Originated as an answer to the problems posed by the ELAIS data reduction, 
the LARI method has evolved into a complete and well-tested system for
ISO-CAM/PHOT data reduction and analysis, especially designed for the
detection of faint sources and the interactive check of detected sources.
Raster observations carried out with ISO-CAM LW detector at 7 and 15
$\mu$m and with ISO-PHOT C100 detector at 90 $\mu$m have been
successfully reduced, while tests are foreseen to apply the method to
other ISO detectors, to ISO-PHOT C200 in particular.

Interactive by its very nature, the method both allows ISO-CAM/PHOT
data reduction at all flux levels from scratch and to check
the quality of any independent data reduction undertaking,
thus leading to extremely reliable and complete source catalogues.
It is thus believed that the LARI method can prove a very efficient tool
in providing the community with long-awaited agreed-upon results from ISO
extragalactic surveys, possibly the ultimate legacy of the ISO mission.
%
%In due time, it is planned to release the developed software to the commmunity
%either as a stand-alone software package or within CIA/PIA.
%
\begin{acknowledgements}
This paper is based on observations with ISO, an ESA project with instruments
funded by ESA Member States (especially the PI countries: France, Germany,
the Netherlands and the United Kingdom) and with the participation of ISAS
and NASA.

The ISOCAM data presented in this paper were analysed using CIA, a joint
development by the ESA Astrophysics Division and the ISOCAM Consortium.
The ISOCAM Consortium is led by the ISOCAM PI, C. Cesarsky.

The ISOPHOT data presented in this paper were reduced using PIA, which is
a joint development by the ESA Astrophysics Division and the ISOPHOT Consortium
with the collaboration of the Infrared Processing and Analysis Center (IPAC).
Contributing ISOPHOT Consortium institutes are DIAS, RAL, AIP, MPIK, and MPIA.

This work was partly supported by the "POE" EC TMR Network Programme
(HPRN-CT-2000-00138).
\end{acknowledgements}

%
%%\appendix
%%
%\section{EXTRA ITEMS}
%%
%If necessary, one or more appendices can be included. To do that, use
%the \verb!\appendix! command, and then just use the normal sectioning
%commands afterwards; they will be numbered with roman letters rather
%than with sequential numbers.
%
\end{document}